%% file: mur_mnras.tex
\documentclass[a4paper,fleqn,usenatbib]{mnras}

\usepackage{newtxtext,newtxmath}

\usepackage[T1]{fontenc}
\usepackage{ae,aecompl}

\usepackage{graphicx,psfig}	
\usepackage{amsmath}	
\usepackage{amssymb}	
\usepackage{graphics}

\usepackage{longtable}
\usepackage[font=small,labelfont=bf,margin=\parindent,tableposition=top]{caption}
\usepackage[usenames,dvipsnames,svgnames,table]{xcolor}

\usepackage{hyperref}
 \definecolor{arancio}{rgb}{1,0.5,0}
 \definecolor{viola}{rgb}{0.7,0,1}
 \definecolor{verde}{rgb}{0.2,0.7,0.7}
\newcommand\kms{\ifmmode {\rm km\ s}^{-1} \else km s$^{-1}$\fi} 
\newcommand\FWHM{\ifmmode {\rm FWHM} \else ${\rm FWHM}$\fi}
\newcommand\Lsun{\ifmmode L_{\odot} \else $L_{\odot}$\fi} 
\newcommand\Hbeta{\ifmmode {\rm H}\beta 
 \else H$\beta$\fi}

\def \apjl {ApJL}
\def \apjs {ApJS}
\def \aap {A\&A}

\title[]{Putting the \textbf{\textit{hadron beam}} scenario for extreme blazars to the test with the Cherenkov Telescope Array}

\author[F. Tavecchio et al.]{
F.\ Tavecchio,$^{1}$\thanks{E-mail:fabrizio.tavecchio@brera.inaf.it}
P.\ Romano,$^{1}$
M.\ Landoni$^{1}$ 
S.\ Vercellone,$^{1}$ \\
$^{1}$INAF, Osservatorio Astronomico di Brera, Via E.\ Bianchi 46, I-23807, Merate, Italy\\
}

\date{In original form \today}

\pubyear{2018}

\begin{document}
\label{firstpage}
\pagerange{\pageref{firstpage}--\pageref{lastpage}}
\maketitle

\begin{abstract}
{\it Hadron beams} are invoked to explain the peculiar properties of a subclass of BL Lac objects, the so-called extreme BL Lacs (EHBLs). This scenario predicts a quite distinctive feature for the high-energy gamma-ray spectrum of these sources, namely a hard energy tail extending up to $\sim100$ TeV. It has been proposed that the detection of this tail can offer an unambiguous way to distinguish between the hadron beam scenario and the standard one, which instead assumes gamma-ray emission from the jet strongly depleted at the highest energies ($E>30$ TeV) because of the interaction with the  optical-IR cosmic radiation field. We present dedicated simulations of observations through the presently under construction Cherenkov Telescope Array (CTA) of the very-high energy spectrum of the prototypical EHBL 1ES 0229+200 assuming the two alternative models. We demonstrate that, considering 50 hours of observations from the southern site of CTA (the most sensitive at the highest energies), in the case of the hadron beam model it is possible to detect the source up to 100 TeV. This, together with the non detection of the source above 10 TeV in the standard case, ensures that CTA observations can be effectively used to unambiguously confirm or rule out the hadron beam scenario. 
\end{abstract}

\begin{keywords}
galaxies: radiation mechanisms: non-thermal -- BL Lacertae objects: general --  BL Lacertae objects: individual: 1ES 0229+200 --  gamma-rays: galaxies
\end{keywords}


%

 	 \section{Introduction \label{murase:intro}} 
%

Blazars (e.g. Romero et al. 2017) constitute the majority of extragalactic $\gamma$-ray sources, both at GeV (e.g., Acero et al. 2015, Massaro et al. 2015) and TeV (e.g., Rieger et al. 2013) energies. Their powerful and variable non-thermal emission spans the entire electromagnetic spectrum, from radio waves to very-high energy (VHE) $\gamma$ rays. The extreme observational properties of blazars are explained as due to the fact that they host a relativistic jet pointing toward the observer (Blandford \& K{\"o}nigl 1979). In this conditions, the relativistic effects are maximized, leading to pronounced beaming and amplification of non-thermal emission produced in the outflowing plasma.

The broad band spectral energy distribution of blazars (SED) displays two characteristic broad ``humps". The low energy component (with maximum at IR-X--ray frequencies, depending on the specific source) derives from synchrotron emission of relativistic electrons. The high-energy component, peaking at $\gamma$-ray energies, is often associated to inverse Compton (IC) emission by the same electron population. For BL Lac objects (which represent the great majority of blazars detected at TeV energies) it is generally assumed that the target soft photons for the IC scattering are the synchrotron photons themselves (synchrotron self-Compton model, SSC, e.g. Tavecchio et al. 1998). Alternatively, hadronic scenarios interpret the high-energy bump (or its high-energy part) as the left-over of electromagnetic cascades initiated by high-energy protons (Mannheim 1993, Muecke et al. 2003, Zech et al. 2017) or as their synchrotron emission (Aharonian 2000). 

A quite interesting alternative approach has been considered for the so-called {\it extreme} high-energy peaked BL Lacs (EHBL, Costamante et al. 2001), a small group of TeV emitting blazars displaying peculiar properties (Bonnoli et al. 2015, Costamante et al. 2018), whose most studied representative is 1ES 0229+200 (Aharonian et al. 2007, Costamante et al. 2018). The TeV spectrum of these sources (up to $\approx$10 TeV), once corrected for the expected absorption of high-energy photons induced by the interaction with the IR-optical extragalactic background light (EBL), is often exceptionally hard (Aharonian et al. 2007). Furthermore, in contrast with the rest of the blazar population, the VHE emission is only moderately variable on timescales of months (although long-term small-amplitude variability, very difficult to detect because of the very low flux, is present at some level, see Aliu et al. 2014, Cologna et al. 2015). The hard VHE spectrum is  very difficult to interpret in the framework of the standard SSC model. In fact, both the reduction of the Klein-Nishina scattering cross-section at high-energies (e.g. Tavecchio et al. 1998) and the expected relevant absorption of gamma rays within the jet necessarily entail soft spectra above few TeV. Several  solutions to this problem have been proposed, including very high values of the minimum Lorentz factor of the emitting electron (Katarzynski et al. 2006, Tavecchio et al. 2009, Costamante et al. 2018), Maxwellian-like electron energy distribution (Lefa et al. 2011), internal absorption in the source (Aharonian et al. 2008), hadronic processes (e.g. Cerruti et al. 2015), IC scattering of the cosmic microwave background in Thomson regime in the large-scale jet (Boettcher et al. 2008). \\
A quite different view attributes the peculiarities of EHBL to the fact that gamma rays are not originally produced in the blazar jet but, instead, are generated in the intergalactic space by escaping ultra-high energy cosmic rays (UHECR) energized  within the jet and subsequently beamed toward the observer. While traveling toward the Earth, UHECR lose energy through photo-meson and pair production (Bethe-Heitler) reactions, triggering the development of electromagnetic cascades in the intergalactic space. Because of the reduced distance, high-energy gamma rays produced by the cascades experience a less severe absorption and can reach the Earth (e.g., Essey \& Kusenko 2010, Essey et al. 2011, Takami et al. 2013). Although the power demand of this model is higher than that predicted by leptonic models (e.g. Razzaque et al. 2012), it can be kept to an acceptable value if the cosmic ray beam is not enlarged too much by intervening magnetic fields (e.g. Murase et al. 2012, Tavecchio 2014).  Because of the reduced absorption associated to gamma rays produced by the traveling hadron beam, a distinctive prediction of this model is that the observed gamma-ray spectrum extends at energies much higher than those allowed by the conventional propagation through the EBL. For sources located at low redshift ($z\lesssim0.3$) the spectra should be characterized by a hard tail above 10 TeV whose detection is considered  the smoking gun of this model (e.g Murase et al. 2012). For sources at larger redshift, characterized by a more severe attenuation, the tail appears at lower energies. A representative example of a high-redshift source is studied in  Acharya et al. (2018), reporting simulations for the hadron beam scenario applied to KUV 00311-1938, at (tentative) $z=0.61$ (Piranomonte et al. 2007; not confirmed by Pita et al. 2012 which found $z>0.506$). For this source the tail would show up at TeV energies.

The energy range and the sensitivity of current Imaging Atmospheric Cherenkov Telescopes (IACTs), together with the low flux generally associated to EHBL at the highest energies, do not allow us to test the existence of the hard tail predicted by the {\it hadron beam} scenario for sources as 1ES 0229+200. However, with the advent of the Cherenkov Telescope Array (CTA; Actis et al. 2011, Acharya et al. 2013, Hofmann 2017), the situation is going to change soon. In particular, a good sensitivity at energies above 10 TeV is foreseen for the southern site, where an array of 70 small-sized telescopes (SST) will be able to operate up to 100 TeV.

In this paper, taking advantage of dedicated spectral simulations, we assess the potentiality of CTA to test the predictions of the hadron beam scenario, focusing to the test case of the prototypical EHBL 1ES0229+200 for which Murase et al. (2012) presented the spectrum predicted  by the hadron cascade model. This extends the study reported in Acharya et al. (2018), focused on the case of KUV 00311-1938, to the possibility to detect the tails predicted to hadron beam models at tens of TeV.

The paper is organized as follows: in \S 2 we discuss the spectral models used for the simulations, in \S 3 we illustrate the simulations set-up and the results and finally in \S 4 we conclude.

\begin{figure}
\vspace*{-0.6 truecm}
\includegraphics*[angle=0,height=9.5cm]{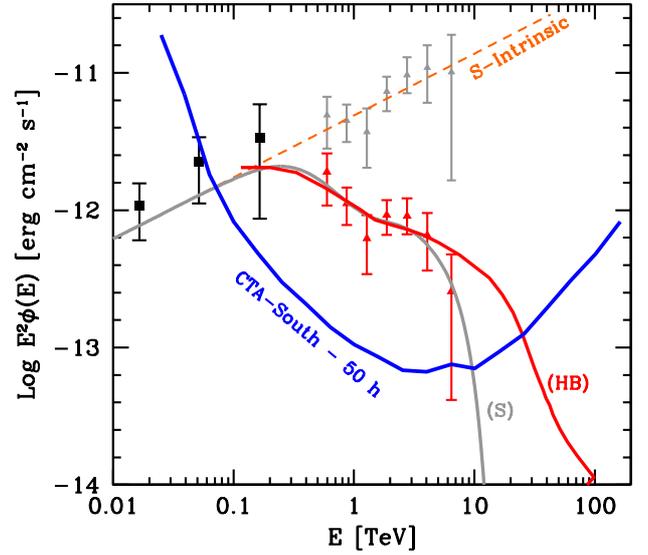}
\vspace{-1 truecm}
\caption{Models for the gamma-ray spectrum of 1ES 0229+200 simulated in the present work.  The data points are from Costamante et al. (2018) (LAT, black squares) and Aharonian et al. (2007) (H.E.S.S., red triangles). The gray triangles show the observed spectral points corrected for the EBL absorption using the model of Dominguez et al. (2011).  The dashed orange line reports the intrinsic spectrum (modelled as a power law $F_E\propto E^{-0.55}$) expected in the case of a standard emission scenario, while the solid gray line show the expected observed spectrum (S). The red solid line reports the model by Murase et al. (2012) for the hadron beam (HB) scenario. For comparison we report the CTA sensitivity foreseen for the South site and 50 hours of observation for objects near zenith (solid blue line).}
\label{fig:models}
\end{figure}

 	 \section{Spectral models \label{murase:model}} 
%
%

The BL Lac object 1ES 0229+200 (z=0.14) is one of the best examples of the EHBL subclass. In particular, its intrinsic VHE spectrum, obtained by correcting the observed one for the  expected  EBL absorption, is extremely hard and it is well reproduced by a hard power law. Fig. \ref{fig:models} shows (red triangles) the H.E.S.S. spectrum reported in Aharonian et al. (2007), together with the points derived by correcting for the EBL absorption with the Dominguez et al. (2011) model (gray). Note that the spectra obtained by VERITAS (Aliu et al. 2014) closely follow that reported by H.E.S.S. We also show the spectrum at GeV energies recently derived by Costamante et al. (2018) analyzing LAT data (black squares). The LAT and the de-absorbed VHE datapoints agree quite well and they describe an unbroken power law up to at least 10 TeV. The orange dashed line shows a phenomenological model for the intrinsic spectrum assuming that the continuum extends with the same slope ($F_E\propto E^{-0.55}$) even above the last H.E.S.S. datapoint. At this level we are not explicitly assuming any specific origin for this emission (i.e. leptonic or hadronic). Taking into account the EBL absorption  we then obtain the gray solid line for the  observed spectrum. This model is expected to describe the spectrum emitted by the source in case of the conventional scenario, in which photons are directly produced in the jet and there is no internal absorption. Note that, because of the very high optical depth, the assumptions on the spectral shape above 10 TeV are relatively unimportant in determining the observed flux.
In the following we will refer to this case as the {\it standard} (S) model.

Murase et al. (2012) calculated the observed gamma-ray spectrum of 1ES 0229+200 in the case of the hadron beam scenario. As discussed above, in this case the gamma rays are produced all along the trajectory of the hadrons from the source to the Earth and therefore the effective absorption is lower than in the standard case. The spectrum (red solid line in Fig.\ref{fig:models}) is almost coincident to the standard one up to several TeV. Clearly, current data are in agreement with both models and cannot be used to distinguish the nature of the gamma-ray emission. However, above 10 TeV the hadron beam scenario predicts a hard tail extending up to 100 TeV and the spectrum 
intersects the CTA sensitivity curve (for 50 hours of observation) at several tens of TeV. On the contrary, the huge absorption implies a severe cut-off of the spectrum for the standard case.
This key difference offers a direct way to test the hadron beam scenario. 
A detection at 20--30 TeV, while ruling out the standard model, will provide a smoking gun for the hadron beam. 

In the next Section we describe the set-up used to perform the simulations of CTA observations for the two alternative models.

 	 \section{Simulations \label{murase:simulations}} 
%
%
%
%

\subsection{Set-up}

For our simulations we used the {\tt ctools}  (Kn{\"o}dlseder et al. 2016)\footnote{\href{http://cta.irap.omp.eu/ctools/}{http://cta.irap.omp.eu/ctools/.} }, 
 analysis package for IACT data,  and the public CTA instrument response 
files\footnote{\href{https://www.cta-observatory.org/science/cta-performance/}{https://www.cta-observatory.org/science/cta-performance/.}  }
(IRF, v.\ prod3b-v1). 
Since the  source is located at RA(J$2000)=38.221667$ deg,  Dec(J$2000)=20.272500$ deg,
it is visible from both CTA sites, at zenith angles of $\sim9^\circ$ from the North and $\sim45^\circ$ from the South.  
Since we selected an exposure of 50 hr we chose the {\tt North\_z20\_average\_50h} and {\tt South\_z40\_average\_50h}  IRFs.Note that  from the South site the zenith angle is always larger then 40 degrees assumed for the latter IRF. However, the present work is focused on the detectability above 10 TeV, a range in which the sensitivity is expected to improve with increasing zenith angle. Therefore our simulation using the IRF valid for 40 degrees can be considered conservative.
A summary of the inputs to our simulations is reported in Table~\ref{murase:table:sims_BL}. 

\begin{figure*}
\hspace{1 truecm}
\includegraphics*[angle=90,height=11.95cm,width=15.2cm]{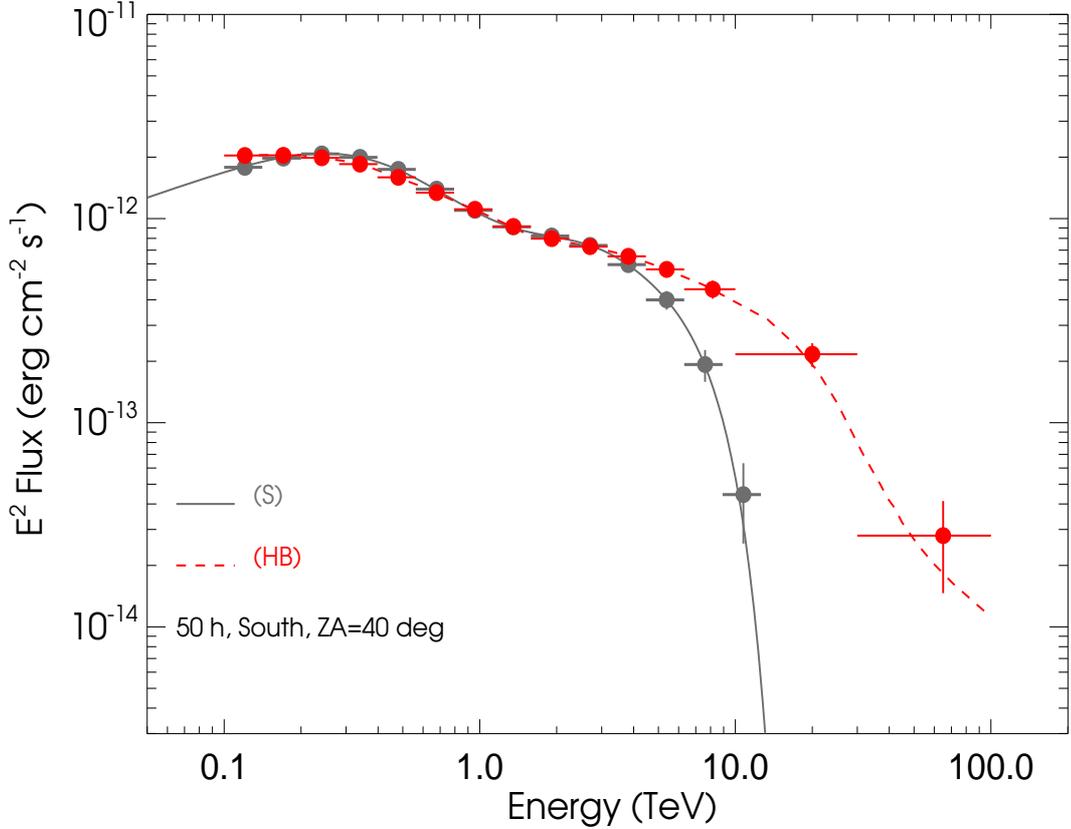}
\caption{Simulated spectra for the standard (gray, southern site) and the hadron beam (red) models using the South effective areas. In the latter case the sensitivity of CTA ensures the significative detection in the high-energy bins (above 10 TeV). See text for details.}
\label{fig:simulations}
\end{figure*}

In the model definition XML file, the spectral model component was introduced as a 
"FileFunction" type, so that the spectrum was provided as an ASCII file containing 
energy (in MeV) and differential flux values (in units of ph\,cm$^{-2}$\,s$^{-1}$\,MeV$^{-1}$), 
described according to 
\begin{equation}
M_{\rm spectral}(E)=N_0 \frac{dN}{dE},
\end{equation} 
where $N_0$ is the normalisation. 
%
For the background we considered only the instrumental one included in the IRFs 
({\tt CTAIrfBackground}) and no further contaminating astrophysical sources in the 
5\,deg field of view is assumed for event extraction. 

\begin{figure}
\includegraphics[angle=0,width=9.3cm]{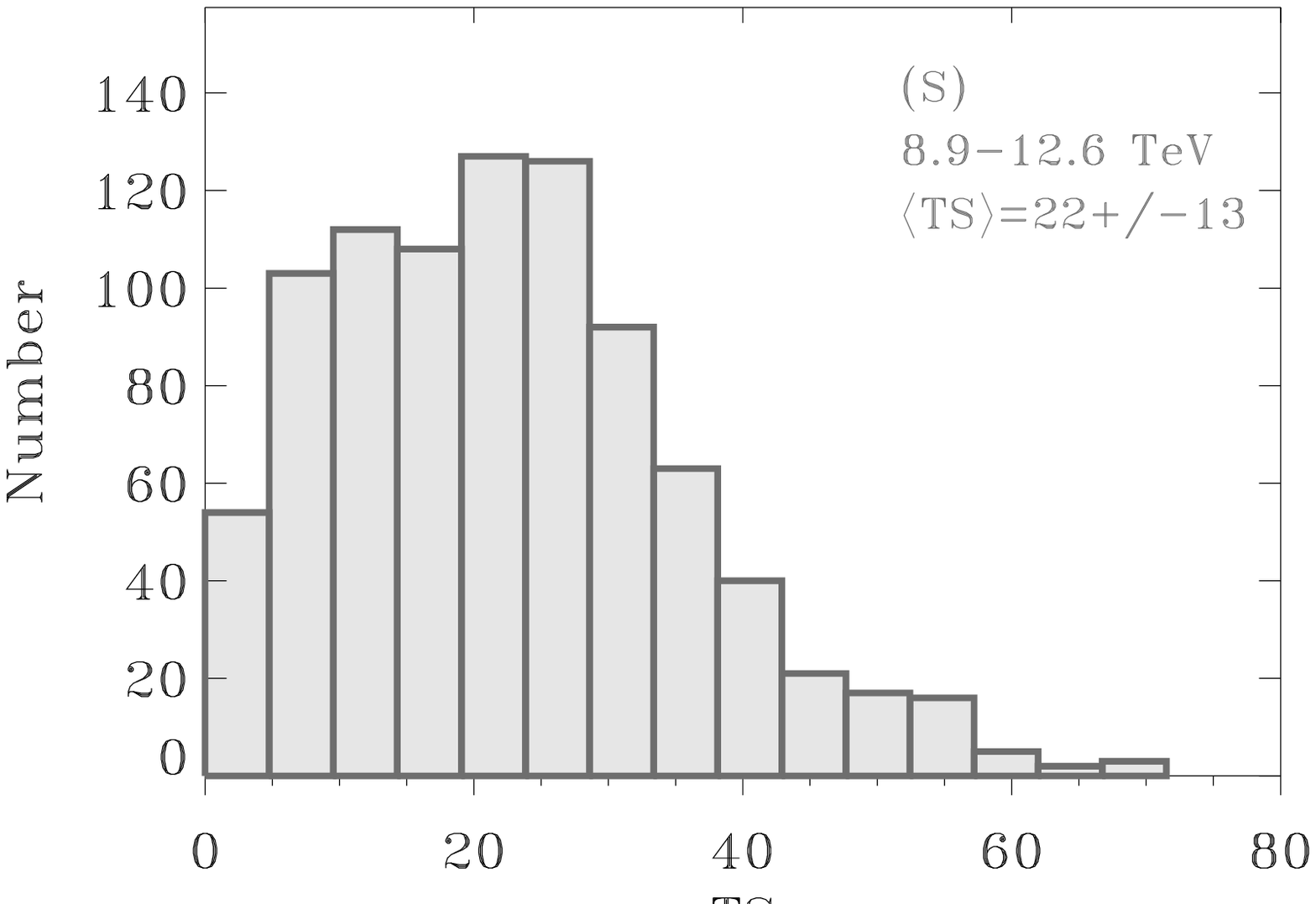}
\caption{TS distribution of the 1000 realizations for the energy bin 8.9-12.6 TeV for the S model (S Site). The average TS value is also reported. The percentage of realizations with $TS>10$ is 81\%.}
\label{fig:histo3}
\end{figure}
\begin{figure}
\hspace{0. truecm}
\includegraphics[angle=0,height=6.cm]{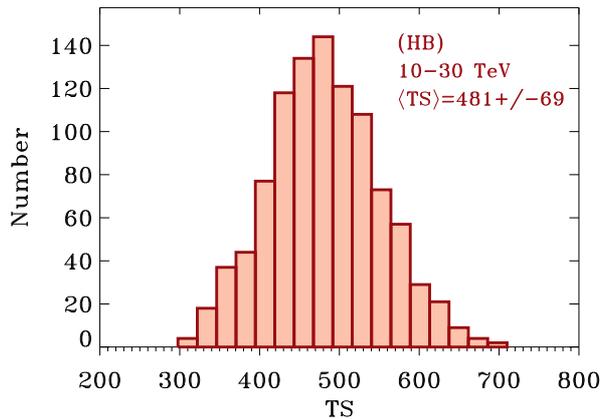}
\caption{TS distribution of the 1000 realizations for the energy bin 10-30 TeV for the HB model. The average TS value is also reported.}
\label{fig:histo1}
\end{figure}
\begin{figure}
\hspace{0. truecm}
\includegraphics[angle=0,height=6cm]{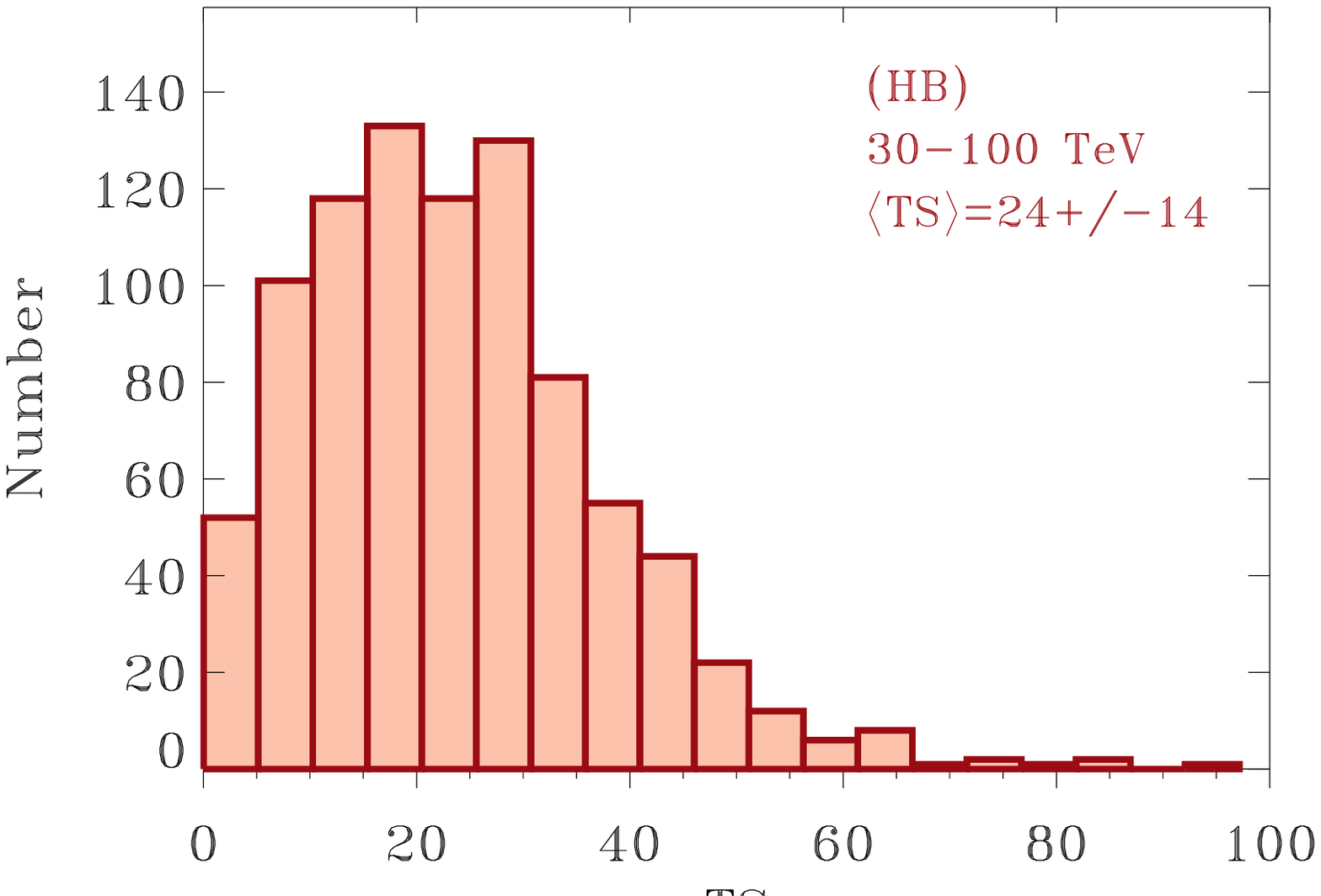}
\caption{TS distribution of the 1000 realizations for the energy bin 30-100 TeV for the HB model. The average TS value is also reported. The percentage of realizations with $TS>10$ is 84\%.}
\label{fig:histo2}
\end{figure}

%
          \input{tab_mur_mnras.tex}


%
We considered a set of 15 energy bins covering an energy range reported in 
Table~\ref{murase:table:sims_BL} (Col.\ 7). We adapt the binning to the the characteristics of the CTA sensitivity and the shape of the hadron beam. In particular we consider larger bins for energies above 10 TeV, where the expected number of recorded events is limited.
In each bin we used the task {\tt ctobssim} to create 
event lists based on our input models, 
including the randomized background events. 
%
We then used the task {\tt ctlike} to fit each spectral bin with a power-law model 
\begin{equation}
M_{\rm spectral}(E)=k_0 \left( \frac{E}{E_0} \right) ^{\gamma},
\end{equation}
where $k_0$ is the normalisation (or {\tt Prefactor}, in units of ph\,cm$^{-2}$\,s$^{-1}$\,MeV$^{-1}$)
$E_0$ is the pivot energy ({\tt PivotEnergy} in MeV), 
and  $\gamma$ is the power-law photon index ({\tt Index}). In the fits we kept 
{\tt PivotEnergy} fixed, at $10^{6}$\,MeV, while {\tt Prefactor} and {\tt Index} were left to vary. 
We thus obtained the spectral parameters and hence the fluxes 
of our gamma ray source in each bin by using maximum likelihood model fitting. 
Statistical uncertainties of the parameters were also calculated, 
as well as the test statistic (TS) value (Cash 1979, Mattox et al. 1996).
For each spectral model we performed sets of $N=1000$ 
statistically independent realisations\footnote{In order  
to efficiently run such large number of simulations,  
we performed them through Amazon Web Services,  
following the methods described in Landoni et al. (2018, in prep). } 
by adopting a different seed for the randomization ({\tt seed}) 
in order to reduce the impact of variations between individual realisations 
(see, e.g. Kn{\"o}dlseder et al. 2016). 
We thus obtained a set of 1000 values of each spectral parameter (and TS)
from which 1000 values of fluxes were calculated in each energy bin. 
For each energy bin, we adopted as flux and error the 
mean and the square root of the standard deviation 
obtained from the distribution of fluxes\footnote{Mean 
flux $\overline{F_{\rm sim}}  = \frac{1}{N}\sum_{k=1}^{N}F_{\rm sim}(k)$, 
standard deviation $s^2_{\rm sim}=\frac{1}{N-1}\sum_{k=1}^{N}(F_{\rm sim}(k)-\overline{F_{\rm sim}})^2$. }.

	 \subsection{Results} 
%

The spectra obtained with the simulations are shown in Fig. \ref{fig:simulations}. In both cases we report the input spectrum (solid line: standard; dashed line: hadron beam) and the reconstructed spectrum in bins of energy.

In the case of the standard model, the simulations clearly indicate that the spectrum can be detected only up to energies around 10 TeV from the southern site. In fact the steep spectrum caused by the absorption implies no signal at the highest energies. Using the northern array, less sensitive at the highest energy, the source is detected only up to about 7 TeV. On the other hand, in the case of the hadron beam spectrum, the hard energy tail can be tracked up to the highest energy bin, close to 100 TeV. The significance of the detection in this last energy bin ($TS=23.8$) ensures a solid detection up to these energies. The comparison with Fig. \ref{fig:models} shows that at these energies the spectrum lies below the CTA sensitivity curves. To this respect it is worth to recall that the sensitivity curve reported in Fig. \ref{fig:models} is derived assuming quite conservative parameters. In particular, it is calculated assuming five logarithmic bins in energy per decade and requires 5-standard-deviation, at least 10 detected gamma rays per bin, and a signal/background ratio of at least 1/20 (Bernl{\"o}hr et al. 2013).  These conditions are clearly relaxed in the derivation of the bins and this explains why detections are also possible for fluxes below the sensitivity curve.

To further demonstrate the CTA power to track the spectrum at the highest energies we report in Figs. \ref{fig:histo3}-\ref{fig:histo2} the distributions of the $TS$ for the 1000 realizations. For the standard model the simulations show a clear detection only up to $\sim 10$ TeV for the southern array (see Fig. \ref{fig:histo3}). On the other hand, Fig. \ref{fig:histo1} clearly show the robustness of the detection in the 10-30 TeV band, the ``smoking gun" of the HB model. Even at the bin at the largest energy (Fig. \ref{fig:histo2}) for the great majority of the realizations the TS ($84\%$) is above 10.  We can conclude that the simulated CTA observation would unambiguously discriminate between the two competing models.


 	 \section{Discussion \label{murase:discussion}} 
%

The nature of the so-called EHBL is still matter of discussion and their peculiarities are not easily understood in the standard framework for blazars (e.g. Costamante et al. 2018). In this context the hadron beam scenario offers a natural explanation for their spectral properties at very high energy. We have shown that with a moderately deep exposure of CTA one could unambiguously test the key prediction of this model for the prototypical EHBL 1ES 0229+200, namely the existence of a hard tail extending up to several tens of TeV. In this task the main players will be the SSTs, thanks to which the sensitivity of CTA will be extended at energies much larger than those reached by current IACTs. 
The very limited variability foreseen for the hadronic hard tail emission (e.g. Prosekin et al. 2012) is compatible with a flexible scheduling of the observations.

In this work we limit the discussion to the case study of 1ES0229+200. Being one of the brightest EHBL at VHE (e.g. Costamante et al. 2018), this source offers the best opportunity to test the prediction of the hadron beam model.  We used the specific spectrum provided by Murase et al. (2012) but the details of the resulting cascade spectrum are expected to be quite insensitive on parameters, such as the slope and the maximum energy of the injected cosmic rays. However, the features displayed by the high-energy tail depends on the redshift of the sources (see cases studied in Takami et al. 2013 and that reported in Acharya et al. 2018). In particular, for sources located at (relatively) large redshift (such as the case of  KUV 00311--1938 studied in Acharya et al. 2018) the EBL absorption is already important at a few hundreds of GeV and the tail produced by the hadron beam appears in the TeV band. Since the performance of CTA at TeV energies (where the dominant telescopes are the medium-sized telescopes) substantially differs from that at the energy range above 10 TeV (probed by SSTs), it is important to investigate the prospects of observations of sources at different redshift. In view of the selection of the best EHBL to be observed by CTA, an effort should be made to find more candidates and to study in a more systematic way the predicted properties of the cascade component.

We worked under the (possibly extreme but simple) hypothesis that the reprocessed cascade emission accounts for the entire gamma-ray component of EHBL. However (as suggested by the variability, although of low amplitude) one cannot exclude that the observed VHE spectrum is the result of the contribution of an hadronic-initiated cascade and a direct photon emission from the jet (e.g. Tavecchio 2014). In that case the flux of the high-energy tail can be lower than that assumed here, the precise value being fixed by the relative ratio between the two components.
In this context it would be interesting in a future work to assess the lowest contribution from the cascade that would allow the detection of the hard tail by CTA. 

Hard tails related to hadronic emission are also expected in scenarios in which hadrons lose their energy within the jet (e.g. Cerruti et al. 2015, Zech et al. 2017). However, being produced at the source redshift, these high-energy components are suppressed by EBL absorption for sources at relatively large distance. Moreover, this component does not show the spectral shape and temporal behavior expected for the hadron beam tails and therefore it can be distinguished from it.

This paper focused on the possibility to test the hadron beam model through the study of the spectrum of EHBL. Another important tool that can be exploited to discriminate between leptonic and hadronic beam models is variability. In fact, the reprocessed emission triggered by hadron beams is expected to be quasi-steady (e.g. Murase et al. 2012, Prosekin et al. 2012). Currently, studies on the variability shown by EHBL are hampered because of their low flux in the VHE band (e.g. Aliu et al. 2014), that prevent the detection of short-term variability. The greatly improved sensitivity of CTA will also pave the way for a better understanding of this aspect.

In this paper we focus on the idea that the EHBL can be understood as sources of hadron beams. However the interest to this sources is not limited to this possibility. In fact, even if their gamma-ray emission derives from a standard photon beam, the extreme hardness of the spectrum can be exploited to perform a test of Lorentz Invariance Violation (LIV, Tavecchio \& Bonnoli 2016), to measure the intergalactic magnetic field (Neronov \& Vovk 2010, Tavecchio et al. 2010, 2011) or to probe the existence of axion-like particles (Galanti et al. in prep), not to mention the implication of their phenomenology for the physics of relativistic jets. In this respect an effort should be made to enlarge the still limited population of EHBL (see e.g. Bonnoli et al. 2015, Tavecchio \& Bonnoli 2015), a first step to better understand their demography and their role in the larger blazar population.



\section*{Acknowledgements}

%
The authors acknowledge contribution from the grant INAF CTA--SKA, 
``Probing particle acceleration and $\gamma$-ray propagation with CTA and its precursors'' (PI F.\ Tavecchio). \\

%
This research has made use of the NASA/IPAC Extragalactic Database (NED) which is operated 
by the Jet Propulsion Laboratory, California Institute of Technology, under contract with the 
National Aeronautics and Space Administration. \\

          We gratefully acknowledge financial support from the agencies and organizations 
          listed here: 
\href{http://www.cta-observatory.org/consortium_acknowledgments.}{http://www.cta-observatory.org/consortium\_acknowledgments.}

%
This research made use of ctools, a community-developed analysis package for Imaging Air Cherenkov Telescope data. 
ctools is based on GammaLib, a community-developed toolbox for the high-level analysis of astronomical gamma-ray data. \\ 

%
This research has made use of the CTA instrument response functions provided by the CTA Consortium and Observatory, 
see \href{https://www.cta-observatory.org/science/cta-performance/}{https://www.cta-observatory.org/science/cta-performance/} 
(version prod3b-v1) for more details. \\ 

%
This paper has gone through internal review by the CTA Consortium. \\ 

%

\bsp	
\label{lastpage}
\end{document}

%% file: tab_mur_mnras.tex
\setcounter{table}{0 } 
 \begin{table*} 
 \begin{center} 	
 \caption{Array of {\tt ctools} simulations.        \label{murase:table:sims_BL}} 	
 
 \begin{tabular}{llllllcl} 
 \hline 
 \noalign{\smallskip} 
 Model	   &	Site	&	ZA	&	IRF	&	Exposure	&	Bins	& Energy	&	Number	\\ 
                   & 	        & 	 (deg)       & 	        & 	  (h)           & 	        & (TeV)	&	                \\ 
 \noalign{\smallskip} 
 \hline 
 \noalign{\smallskip} 
   Standard (S)	&	N	&	20	&	{\tt North\_z20\_average\_50h}	        &	50	&	15	& 0.1--30	&	1000	\\ 
   Standard (S)	&	S	&	40	&	{\tt South\_z40\_average\_50h}	        &	50	&	15	& 0.1--30	&	1000	\\ 
 \noalign{\smallskip} 
   Hadron beam	 (HB)        &	S	&	40	&	{\tt South\_z40\_average\_50h}	        &	50	&	15	& 0.1--100	&	1000	\\ 
 \noalign{\smallskip} 
   \hline
  \end{tabular}
 \end{center}
 \begin{list}{}{}
\item 
 \item {(S): Standard=photon emission +EBL absorption; (HB): Hadron beam. ZA: zenith angle.} 
 \end{list}
  \end{table*}